# Some Insights in Superdiffusive Transport

## M. Marseguerra [1], A. Zoia[2]

*Department of Nuclear Engineering, Polytechnic of Milan, Via Ponzio 34/3, 20133 Milan, Italy*

## Abstract

In a wide range of systems, the relaxation in response to an initial pulse has been experimentally found to follow a nonlinear relationship for the mean squared displacement, of the kind $\langle x^2(t) \rangle \propto t^\alpha$, where $\alpha$ may be greater or smaller than *1*. Such phenomena have been described under the generic term of anomalous diffusion. "Lévy flights" stochastic processes lead to superdiffusive behavior ($1 < \alpha < 2$) and have been recently proposed to model – among the others – the subsurface contaminant spread in highly heterogeneous media under the effects of water flow. In this paper, within the Continuous-Time Random Walk (CTRW) approach to anomalous diffusion, we compare the analytical solution of the approximated Fractional Diffusion Equation (FDE) with the Monte Carlo one, obtained by simulating the superdiffusive behavior of an ensemble of particle in a medium. We show that the two are neatly different as the process approaches the standard diffusive behavior. We argue that this is due to a truncation in the Fourier space expansion introduced by the FDE approach. We propose a second order correction to this expansion and numerically solve the CTRW model under this hypothesis: the accuracy of the results thus obtained is validated through Monte Carlo simulation over all the superdiffusive range. The same kind of discrepancy is shown to occur also in the derivation of the fractional moments of the distribution: analogous corrections are proposed and validated through the Monte Carlo approach.

**1. Introduction**

A general approach to the analysis of transport phenomena is based on Continuous-Time Random Walk (CTRW) [1-4], in which the travel of a particle (a walker) in a medium is modelled as a series of jumps of random lengths, separated by random waiting times. The theory of CTRW with algebraically decaying probability distribution functions (pdf's) has been originally introduced in Physics in a series of seminal papers by Weiss, Scher, Montroll and co-workers [4-6] in the late 1960s to explain evidences of anomalous diffusion occurring in the drift-diffusion processes in amorphous semiconductors. The diffusion is called "anomalous" if the mean squared displacement (MSD) is not linearly proportional to time *t* as in the standard Fickian case, but to powers of *t* larger (superdiffusion) or smaller (subdiffusion) than unity. More recently, anomalous diffusion has turned out to be quite ubiquitous in almost every field of science (see e.g. [1-2] for a detailed review) and the CTRW model has been applied with success to interpret the experimental results and to make predictions on the evolution of the examined systems. Such applications concern among the others e.g. morphogen gradient formation in complex biological environments [7], chaotic Hamiltonian systems (with application to the transport of charged particles in turbulent plasma) [8], transport in disordered systems [9], evolution of financial markets [10], dynamics of ad-atoms on the surface of a solid [11] or transport of contaminant particles in groundwater under the combined effect of rock fractures and porosity [12]. Much attention has been paid to this last topic, since many laboratory-scale experiments as well as direct field measurements have evidenced subdiffusive and superdiffusive behaviours of migrating particles. In the subdiffusive case [3,12-15], an algebraically decaying distribution is assumed for the waiting times of the particles in the surrounding medium (instead of the traditional exponential one). This physically means that the particles will have a non-vanishing probability of extremely long sojourn times in the visited locations. The macroscopic effect is that the variance of the particles ensemble grows sublinearly in time (subdiffusion).

In the superdiffusive case, an algebraically decaying pdf is assumed for the jump lengths (instead of the usual Gaussian distribution) and the macroscopic effect is a non vanishing probability of very long jumps (so-called "Lévy flights"): correspondingly, the variance of the particles ensemble grows superlinearly in time [1-2,16-17]. Evidences of superdiffusive migration for very slowly decaying radioactive species were experimentally found in a site of Los Alamos National Laboratory and more recently in the Yucca Mountain project for a high-level radioactive waste repository, where plutonium and other actinides had migrated up to some meters (instead of the few millimeters expected), thanks to the support of colloids [18]. The fractal nature of radioactive contaminants migration due to morphological environmental complexity, which may give rise to anomalous diffusion, has been studied at length e.g. for the nuclear test site of Semipalatinsk [19-21]. If the migration of radioactive and/or toxic particles leaking from a repository is superdiffusive, an analysis performed within a standard (i.e. Fickian) advection-dispersion scheme may lead to severe errors due to the underestimation of the effective contaminant spread after a given time.

---

[1] *Corresponding Author. Tel: +39 02 2399 6355 . Fax: +39 02 2399 6309*
*Email address: marzio.marseguerra@polimi.it (Marzio Marseguerra)*

[2] *Email address: andrea.zoia@polimi.it (Andrea Zoia)*

The analytical approach to CTRW is based on the Chapman-Kolmogorov master equation and consists in considering the asymptotic behaviour of the jump lengths and waiting times distributions of the walkers' travels (the so-called "diffusion limit": see e.g. [1-3]). Within this approach, one resorts to the Laplace and Fourier transformed domain [1] and considers an expansion truncated to the first non-constant term for both the spatial and temporal distributions [16,3,1]. These approximations allow to obtain the elegant Fractional Diffusion Equation (FDE), whose solution can be formally expressed in closed form through the Fox function [22-24,1-2]. However, when the approximation is limited to the first non-constant term (see e.g. [1,3]) and the characteristic exponent of the jump lengths distribution approaches its superior limit (the standard Gaussian diffusive behaviour), the solution becomes less and less acceptable. In this paper we generalize the superdiffusive transport by approximating to the second order the asymptotic expansion of the algebraically decaying spatial distribution. In order to evaluate and compare the degrees of approximation of the FDE approach and the one of our generalization, we perform a Monte Carlo simulation of the walker's travel, making use of the exact space and time distributions. An analogous situation and therefore the necessity of improving the approximation has been demonstrated by Margolin and Berkowitz [3] for subdiffusion in the case of biased transport. Here we will resort to Monte Carlo simulation in order to get accurate solutions to the CTRW equation and to compare them with the FDE: however, it should be remarked that it is also possible to obtain full solutions of CTRW formulation by resorting to an (analytical) partial differential equation form in the Laplace space and then to a numerical Laplace inverse transform [16,25-27].

The paper is organized as follows: in Section 2 we briefly discuss the CTRW approach and the hypotheses under which the FDE approximated model can be derived. In Section 3 we introduce the Monte Carlo approach as a solution of the CTRW model and show that the Monte Carlo curves neatly differ form those of the FDE when the characteristic exponent of the jump lengths distribution is large. In Section 4 we derive a modified Fourier transform expansion for the jump lengths distribution and compute the numerical solution of the CTRW under this assumption. The accuracy of the proposed correction is validated through Monte Carlo simulation. Section 5 is devoted to the so-called fractional moments and again the analytical approximated results and the proposed numerical corrections are compared to the Monte Carlo findings. Conclusions are finally drawn in Section 6. An Appendix is devoted to the details of the required calculations.

## 2. CTRW and FDE approaches to superdiffusion

Let $X(t)$ be a stochastic process describing the motion of a tracer particle (a walker) performing random jumps separated by random waiting times. The associated pdf $P(x,t)$ represents the probability density of the walker being in $X=x$ at time $t$ and it is also called the propagator of the process. The CTRW approach to the description of this stochastic process is based on the probability balance expressed by the Chapman-Kolmogorov integral equation (the so-called Master Equation [1,16,27]), which entails the pdf $P(x,t)$ [1,4]. It can be shown [22,24,1-2,16] that, if $\lambda(x)$ and $w(t)$ are the distributions of the single jump lengths and waiting times, respectively, the Laplace and Fourier transformed expression of $P(x,t)$ satisfies the simple algebraic relation

$$P(k,u) = \frac{1-w(u)}{u}\frac{1}{1-w(u)\lambda(k)} \qquad (1)$$

in the case of a Cauchy problem with initial conditions $\delta(x)\delta(t)$. For convenience, adopting a well-established convention (see e.g. [1-3]), we denote the Fourier or Laplace transform of a pdf by its argument. According to the choice of the waiting times and of the jump lengths distributions, the CTRW approach may give rise to (normal) diffusive, subdiffusive and superdiffusive behaviour of the walker.

In order to model superdiffusion, we adopt here the so-called "Lévy flights" approach [16-17,1], in which the particles trajectories in the phase space are described by a Markovian stochastic process as follows. The waiting times pdf, $w(t)$, is assumed to be any finite-mean distribution, e.g. an exponential pdf with mean $\tau$, so that its Laplace transform in the variable $u$ is (see e.g. [1])

$$w(u) = \frac{1}{1+u\tau} \approx 1 - u\tau. \tag{2}$$

The above approximation has been introduced by considering the behaviour of the particles for long times ($u\tau \ll 1 \leftrightarrow t \gg \tau$).

As for the distribution of the jump lengths, the Central Limit Theorem ensures that any finite-variance pdf gives rise to normal diffusion. A possibility to get out of the diffusive behaviour, yet preserving stability, consists in resorting to an infinite-variance algebraically decaying (power law) distribution of the kind

$$\lambda(x) \approx A|x|^{-\beta-1} \tag{3}$$

where $|x| \to \infty$ and $1 < \beta < 2$. Here $A$ is a normalization constant and $\beta$ is the so-called characteristic exponent of the pdf. Particles performing infinite-variance jumps sampled from (3) separated by finite-mean waiting times are said to perform "Lévy flights", which unavoidably lead to superdiffusion for the particle ensemble [16-17]. In this description, the pdf of flight lengths and waiting times are decoupled and Markovianity of the stochastic process is ensured by the fact that $\langle t \rangle < +\infty$ [10,16]. It can be shown that the pdf (3) obeys the generalized Lévy-Gnedenko's Theorem [23,1,28] and asymptotically converges to a Lévy stable law [1,23]: this means that, analogously as for the Central Limit Theorem, the sum of many successive jump lengths has the same distribution as that of a single jump, up to a scaling factor.

Assuming that the distribution (3) is defined in $|x| \geq \sigma$, where $\sigma$ is a customary "characteristic length", and normalizing, we get:

$$\lambda(x) = \frac{\beta}{2}|x|^{-\beta-1}\sigma^{\beta} \tag{4}$$

whose Fourier transform is

$$\lambda(k) = \int_{-\infty}^{+\infty} e^{-ikx}\lambda(x)\,dx = \frac{\beta}{2}\sigma^{\beta}\int_{-\infty}^{+\infty} e^{-ikx}|x|^{-1-\beta}\,dx. \tag{5}$$

In the "diffusion limit", i.e. for $x \gg \sigma$ or equivalently $k\sigma \ll 1$, we can approximate the above integral. Up to the first non-constant term, the expansion reads:

$$\lambda(k) \approx 1 - c_{\beta}|k\sigma|^{\beta} \tag{6}$$

where $c_{\beta} = \Gamma(1-\beta)\cos(\frac{\pi}{2}\beta)$.

We recognize that expression (6) takes the same form as the first-order expansion of a Lévy stable law characteristic function. Since this approximation constitutes the basic assumption of the Fractional Diffusion Equation (FDE) approach, we can take advantage of the well-known FDE formalism in order to obtain the corresponding elegant closed-form solution. To this aim, the propagator $P(k,u)$ (1) is at first analytically Laplace inverted, to yield the exponential form

$$P(k,t) = \frac{1}{\lambda(k)}e^{-\frac{t}{\tau}\frac{1-\lambda(k)}{\lambda(k)}} \tag{7}$$

which in the "diffusion limit" can be approximated as

$$P(k,t) \approx e^{-\frac{t}{\tau}(1-\lambda(k))} \tag{8}$$

To finally obtain the desired solution $P(x,t)$, we must perform the Fourier inverse transform of $P(k,t)$. Within the FDE approach to anomalous diffusion, i.e. by substituting expression (6) into equation (8), $P(k,t)$ becomes formally the characteristic function of a Lévy stable distribution, which admits an explicit Fourier inverse transform in terms of the Fox's $H$ function representation [22-24,1-2]:

$$P(x,t) = \frac{1}{\beta x} H_{2,2}^{1,1}\left[\frac{|x|}{(D_\beta t)^{\frac{1}{\beta}}} \middle| \begin{array}{cc} \left(1,\frac{1}{\beta}\right) & \left(1,\frac{1}{2}\right) \\ (1,1) & \left(1,\frac{1}{2}\right) \end{array}\right], \qquad (9)$$

where $D_\beta = \frac{\sigma^\beta}{\tau} c_\beta$ is the generalized diffusion coefficient. Equation (9) is the formal solution of the Fractional Diffusion Equation

$$\frac{\partial}{\partial t} P(x,t) = D_\beta \partial_{-\infty,x}^\beta P(x,t), \qquad (10)$$

where $\partial_{-\infty,x}^\beta$ is the (symmetric) Weyl fractional differential operator [1,24,22]. As a general consideration, equation (9) represents the exact solution of the CTRW model under the hypothesis of approximating $w(u)$ and $\lambda(k)$ with Eqs. (2) and (6), respectively. In the following we shall call (9) the FDE solution of (1). A few words are due on the on the distinct merits and the potential limits of the fractional-in-space derivative model (10). As for its drawbacks, FDE arises as an approximation of a more general and exact transport model, formulated in terms of the Chapman-Kolmogorov Master Equation, as shown in this Section. Moreover, while anomalous diffusion is often experimentally found to be a transient phase, which – after a suitable time interval – generally relaxes towards Fickian diffusion, FDE can not to take into account this transition in a straightforward manner, since the anomalous behaviour is assumed to hold even at infinite time [16,1-2,25-27]. On the other hand, a prominent advantage of FDE with respect to the CTRW approach is that the fractional derivative formulation may easily include external fields in a simple manner and it is naturally suitable to solve boundary value problems [29-31,1-2]. In this respect, FDE has been recently shown to act as a unifying framework for the quantitative description of different physical phenomena where anomalous diffusion plays a significant role [1-2]. Moreover, a plethora of standard mathematical techniques derived from partial differential equations literature are readily available to obtain analytical solutions for FDE. For a broader comparison between the FDE and the classical Fickian advection-dispersion equation as applied to model contaminant subsurface transport we refer the reader e.g. to [32-34]: in particular, Lévy flights significance as applied to model physically realistic transport phenomena has been questioned, since they do not possess a finite second moment, whereas massive particles can not perform extremely long jumps instantaneously [1,16].

## 3. The Monte Carlo approach

In order to evaluate the effects of the approximation (6) in the Fourier space on the final functional form of the propagator $P(x,t)$ in the direct space, we have resorted to the Monte Carlo (MC) simulation. The MC approach to the particle transport is based on the simulation of the *microscopic* behaviour of a large number of particles, evolving in phase space according to the pdf's in the Chapman-Kolmogorov master equation and *macroscopically* resulting in the behaviour of the ensemble. In the case of the anomalous transport, a distinct merit of the MC approach with respect to the Fractional Diffusion Equation is that the simulation is based on the exact $w(t)$ and $\lambda(x)$ pdf's governing the fate of each particle, without the necessity of resorting to the various approximations detailed in the preceding Section. Therefore, the MC simulation may be thought as an exact solution of the CTRW model and therefore utilized to quantitatively evaluate the accuracy of the approximate analytical FDE results. This approach has been proposed e.g. in [26] in order to evaluate the accuracy of full solutions to CTRW formulation as obtained through numerical Laplace inverse transform. In a computer code, before starting the stochastic simulation, the extension of the phase space available for the walker, namely the viable space $(x_{\min}, x_{\max})$ and time $t_{\max}$, must be fixed. Then, this rectangular domain is discretized by assigning the respective numbers of subintervals $N_x$ and $N_t$: correspondingly a matrix $P$ of order $(N_x, N_t)$ is determined, whose generic element $P(i,j)$ at the end of the simulation will contain the number of passages of the walker in the elementary $ij$-th cell. To avoid distortions in the pdf $P$ and then gross errors in the distribution moments as a function of time, it is important that the space assignment be done so that essentially no walks will exit from the space interval before $t_{\max}$.

The simulation of the fate of each superdiffusive particle is performed by sampling the successive jump lengths $\Delta x$ and waiting times $\Delta t$ from the exact pdf's $\lambda(x) = \frac{\beta}{2}|x|^{-\beta-1}\sigma^\beta$ and $w(t) = \frac{1}{\tau}e^{-\frac{t}{\tau}}$, respectively. Specifically, at the generic step of the simulation, the walker arrives at a point $(x,t)$ within the cell $(i,j)$ and the new $\Delta x$ and $\Delta t$ are successively sampled. This implies that the walker remains in the $i$-th spatial interval until $t'=t+\Delta t$ and then it suddenly performs a spatial jump towards the new point $(x' = x + \Delta x, t')$ within the cell $(i + n_i, j + m_j)$ with $n_i$ and $m_j$ integers. Then, a unit is added to all the elements $(i, j), (i+1, j), \ldots (i+(n_i -1), j)$ of the matrix $P$ and the walk continues from $(x',t')$. When a large number of stories have been processed in this way and the walker passages through the discrete cells have been registered, the space-time matrix $P(i,j)$, normalized column by column, will represent the (discretized) distribution of the particles: in particular, the $j$-th column represents the spatial distribution at time $j$.

In the following Figures we show the Monte Carlo simulation results in the case of exponential waiting times with mean $\tau$ and jump pdf (4) with $\beta=1.3$ and $\beta=1.9$ respectively, with $10^5$ processed stories. We show the MC distribution $P(x,t)$ as a function of $x$ when $t=90$. The simulation parameters are $\tau=1$ and $\sigma=10^{-3}$. We remark that when $\beta$ is small (Figure 1) the Monte Carlo is in nice agreement with the FDE solution (9). This means that in this case the FDE yields a good solution to the superdiffusive transport problem. When on the contrary $\beta$ is large (Figure 2), the Monte Carlo curve neatly differs from the analytical FDE solution. In this latter case, the FDE does not yield a good solution and we should conclude that the effects of the approximation leading to equation (6) are not negligible. In practice, the FDE model gives good results only up to approximately $\beta \leq 1.5$.

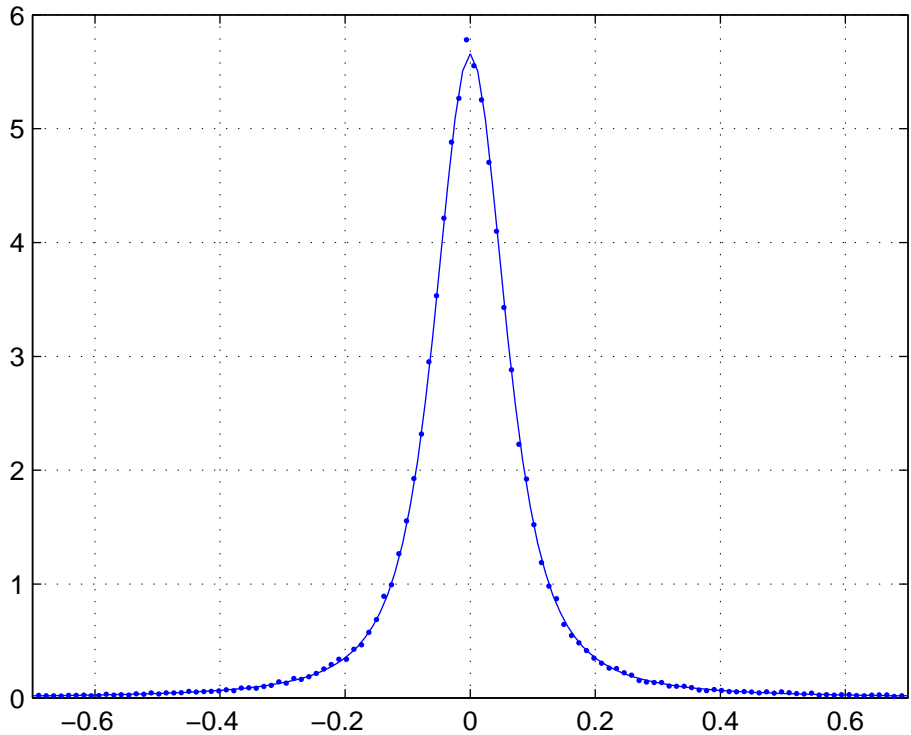

*Figure 1. $\beta=1.3$. Monte Carlo pdf (dots) and FDE analytical solution (solid line).*

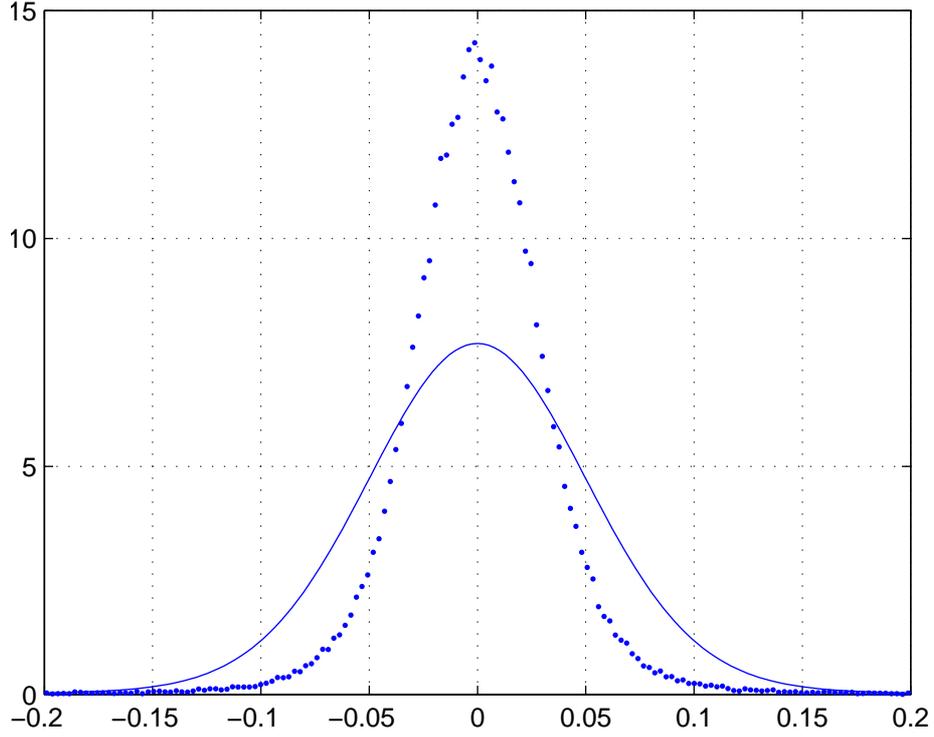

*Figure 2. β=1.9. Monte Carlo pdf (dots) and FDE analytical solution (solid line).*

## 4. The Fourier transform of the jump lengths distribution

Motivated by the limitations arisen in comparing the FDE solution and the MC simulation in the previous Section, we extend to the second order the expansion of the Fourier transform of pdf (4) for the case $1 < \beta < 2$, aiming at improving the accuracy of the solution when $\beta$ approaches its superior limit *2*. The details of the calculations are reported in the Appendix. The final Fourier expansion for the jump lengths distribution reads:

$$\lambda(k) = 1 - c_\beta (|k|\sigma)^\beta + c_2 (|k|\sigma)^2 + o(k^4) \qquad (11)$$

where $c_\beta = \Gamma(1-\beta)\cos(\frac{\pi}{2}\beta)$ and $c_2 = \dfrac{\beta}{2(2-\beta)}$. (12)

We remark that, in addition to the leading term $|k|^\beta$, which is present also in the approximated transform (6), in equation (11) a term quadratic in *k* appears, whose importance is expected to grow as $\beta \to 2$. A first idea of the difference between dropping or maintaining the $|k|^2$ term in the expression (11) of $\lambda(k)$ may be obtained directly from Figures 3a and 3b, which give the plot of the two corresponding propagators *P(k,t)* vs. *k* (at a given time *t*) for a small and a large value of $\beta$ (assuming equal σ for both cases). The figures confirm that for small $\beta$ the $|k|^2$ term is inessential, while it might be important as $\beta$ approaches *2*.

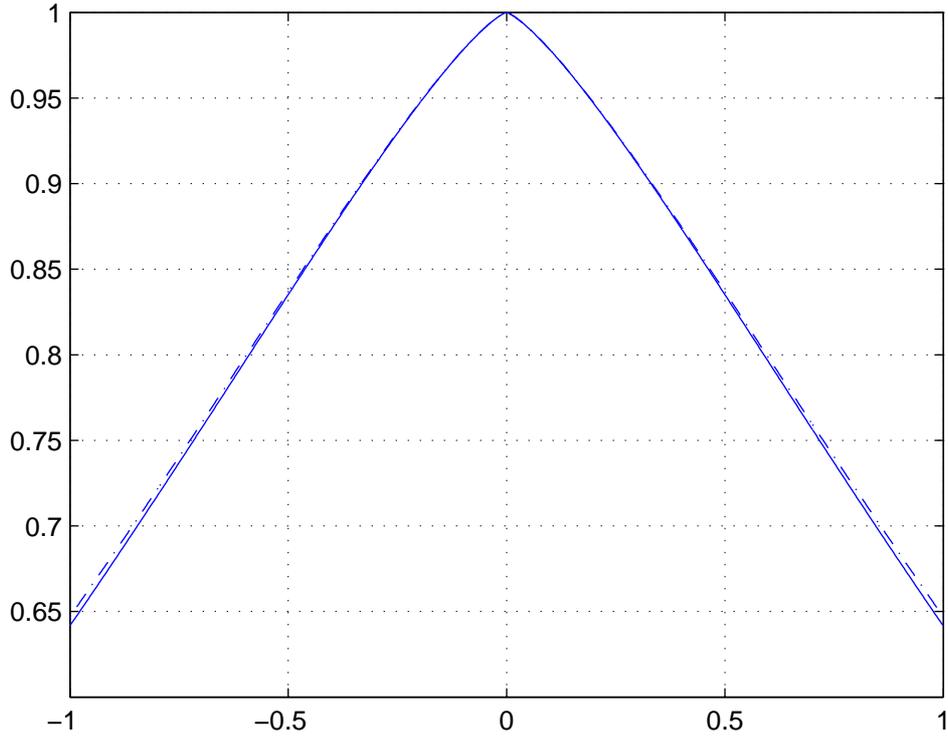

*Figure 3a. The propagator P(k,t) in Fourier k-space (at fixed t) for β=1.3. Solid line: second order expansion (11). Dashed line: FDE approach (6) which neglects the quadratic contribution.*

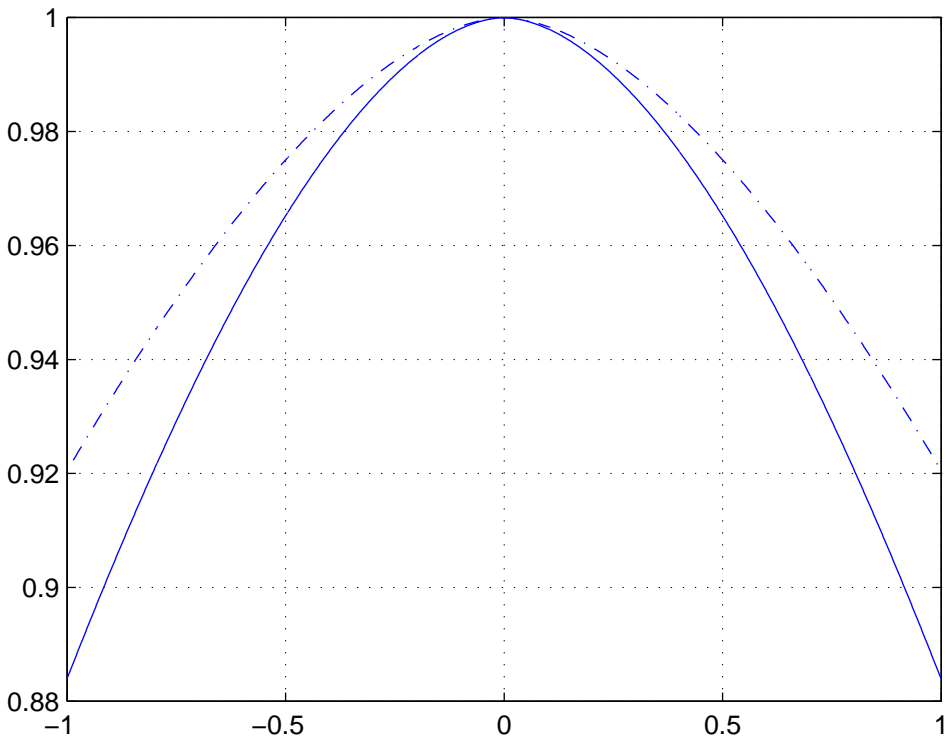

*Figure 3b. The propagator P(k,t) in Fourier k-space (at fixed t) for β=1.8. Solid line: second order expansion (11). Dashed line: FDE approach (6) which neglects the quadratic contribution.*

The presence of the quadratic term in the Fourier transform justifies the fact that the solution (9) of FDE model, which follows from the expansion of $\lambda(k)$ to the first non constant term [1,16], is an increasingly worse approximation of the exact solution (i.e. the MC distribution) as $\beta \to 2$. This situation is in facts very similar to the one occurring in the subdiffusive case, where the explicit Laplace expansion of the waiting times distribution (which is a power law pdf with characteristic exponent $\alpha$, with $0 < \alpha < 1$) contains a linear term in the transformed variable $u$ together with the dominant contribution of order $u^\alpha$. In this latter case, the importance of the linear term grows as $\alpha \to 1$ and correspondingly the FDE yields a solution less and less satisfactory with respect to the one of the MC approach [3].

When $\lambda(k)$ is assumed of the kind (11), the propagator $P(k,t)$ has no known analytical closed-form Fourier inverse. However, the solution $P(x,t)$ may be numerically computed by resorting to an IFFT algorithm for the inversion of (8). In the following Figures we compare the hybrid solution so obtained with the Monte Carlo one, which – according to the previous considerations – may be thought as the reference curve. We observe that both for small (Figure 4) and large $\beta$ (Figure 5) the hybrid solution, computed including the quadratic term in the Fourier transform of the pdf (4) is in remarkably good agreement with the Monte Carlo[3]. This allows us to conclude that in the "diffusion limit" (which is justified by our choice of parameters, namely a small $\sigma$ and times much larger that $\tau$) it is not necessary to include other terms beyond the second order in the Fourier expansion (11).

We remark that the discrepancy between the Monte Carlo and hybrid solution on one side and the FDE on the other side becomes negligible as $x >> \sigma$ (see e.g. remarks in [1,16]). However, when $\beta$ is large – say $\beta > 1.5$ – this condition can be satisfied only for extremely small values of $\sigma$ or, equivalently, for extremely large values of $x$. In this sense, the hybrid solution extends the limits of applicability of the FDE theory to interpret experimental evidences of superdiffusive transport.

In this respect, it is interesting to observe that – analogously as equation (10) can be derived from (1) by exploiting the formal properties of the Fourier and Laplace in terms of (pseudo-)differential operators, under the hypothesis of truncating (5) to the first non-constant term – when expansion (6) is replaced by (11) the same formal properties lead to a generalized FDE, which reads:

$$\frac{\partial}{\partial t} P(x,t) = D_\beta \partial^\beta_{-\infty,x} P(x,t) - D_2 \frac{\partial^2}{\partial x^2} P(x,t) \qquad (10^*)$$

where $D_2 = c_2 \frac{\sigma^2}{\tau}$. We can interpret this expression by stating that, when $\beta$ is small, superdiffusion (i.e. the operator $D_\beta \partial^\beta_{-\infty,x}$) is the dominant process and equation (10*) is almost indistinguishable from (10). However, as $\beta$ grows closer to 2, superdiffusion begins to compete with a usual Fickian diffusion process, which is condensed in the differential operator $D_2 \frac{\partial^2}{\partial x^2}$. According to the previous considerations, it follows that the Monte Carlo and hybrid solutions formally satisfy equation (10*).

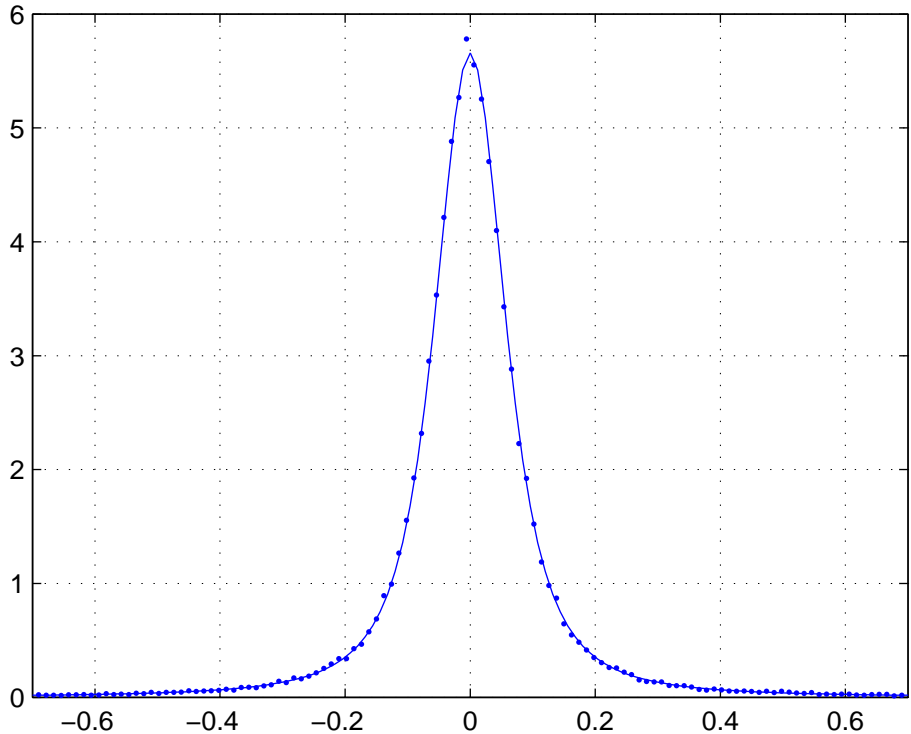

*Figure 4. $\beta=1.3$. Monte Carlo pdf (dots) and hybrid CTRW solution (solid line).*

---

[3] The simulation parameters are the same as in the previous Section and the distribution $P(x,t)$ is plotted vs. $x$ for $t=90$.

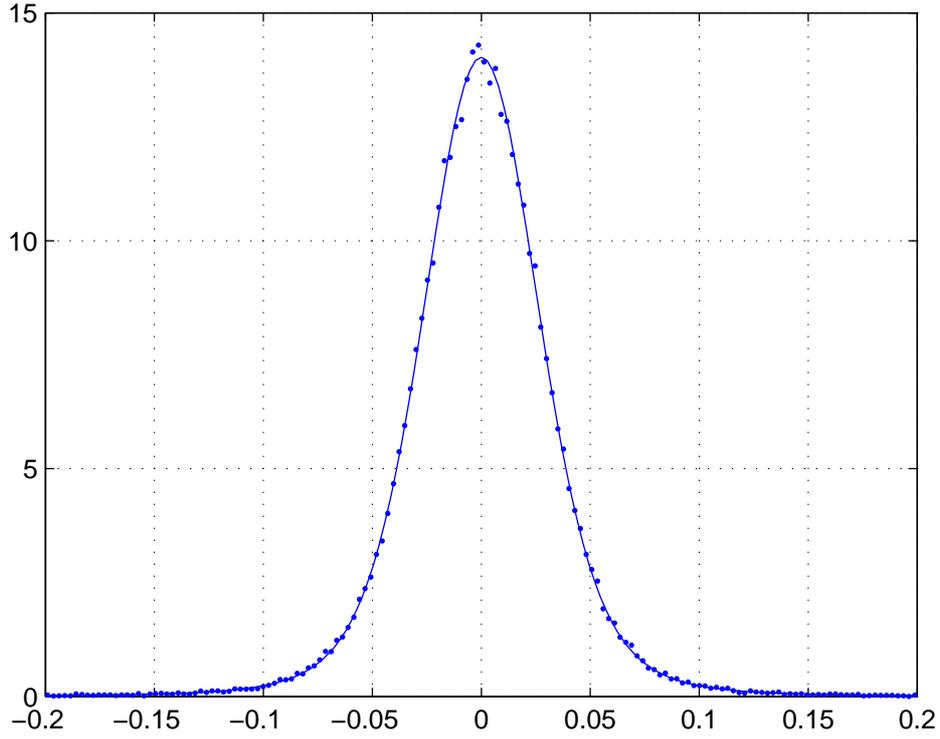

*Figure 5. $\beta=1.9$. Monte Carlo pdf (dots) and hybrid CTRW solution (solid line).*

## 5. The fractional moments

We now extend the considerations of the previous Section to the moments of distribution *P(x,t)*. It is known that the mean squared displacement of a Lévy stable distribution with exponent $1 < \beta < 2$ diverges [1]. However, it is possible to define a *fractional moment* of order $\delta$, as

$$\left\langle |x(t)|^{\delta} \right\rangle = 2 \int_0^{+\infty} x^{\delta} P(x,t)\, dx \tag{13}$$

which exists for $0 < \delta < \beta \leq 2$ [35-36, 1]. The rescaling of such non-integer order moments leads to a pseudo-variance which grows in time as $\left\langle x^2(t) \right\rangle \propto t^{\frac{2}{\beta}}$, hence superdiffusion for $1 < \beta < 2$ [35-37,1]. Now, if we assume for the jump lengths distribution the expression (4) and for its Fourier transform the truncated expansion $\lambda(k) \approx 1 - c_\beta |k\sigma|^\beta$, which leads to the FDE equation, it can be shown that the fractional moments (13) have a simple analytical expression, which is obtained by exploiting the properties of the Fox function (9) and its integral representation in terms of a Mellin transform [35-37,1]:

$$\left\langle |x(t)|^{\delta} \right\rangle = 2\int_0^{+\infty} x^{\delta} \frac{1}{\beta x} H_{2,2}^{1,1}\left[ \frac{|x|}{(D_\beta t)^{\frac{1}{\beta}}} \middle| \begin{array}{cc} \left(1,\frac{1}{\beta}\right) & \left(1,\frac{1}{2}\right) \\ (1,1) & \left(1,\frac{1}{2}\right) \end{array} \right] dx = \frac{2}{\beta} (D_\beta t)^{\frac{\delta}{\beta}} \frac{\Gamma(-\frac{\delta}{\beta})}{\Gamma(-\frac{\delta}{2})} \frac{\Gamma(1+\delta)}{\Gamma(1+\frac{\delta}{2})} \tag{14}$$

On the basis of the considerations exposed in the previous Section, we expect equation (14) be valid when $\beta$ is small, say approximately for $\beta < 1.5$, i.e. within the range of applicability of the FDE equation (10). For the general case $1 < \beta < 2$, we can generalize expression (11) as follows:

$$\left\langle \left|x(t)\right|^{\delta}\right\rangle = 2\int_{0}^{+\infty} x^{\delta}\,\mathfrak{I}^{-1}\left\{e^{-\frac{t}{\tau}(1-\lambda(k))}\right\}dx \tag{15}$$

The integrals appearing in equation (15) can be numerically evaluated. Again, these results may be validated through Monte Carlo simulations, by calculating the matrix $P(i,j)$ mentioned in Section 3 and then deriving the moments of order $\delta$. The simulation parameters and the number of stories are the same as in Section 3. In the following Figures we show the Monte Carlo evaluation of the fractional moments compared to the analytical formula (14) and to the numerically computed curve (15). The fractional moments are plotted vs. time. We remark that when $\beta$ is small ($\beta=1.5$, Figure 6) equation (15) and (14) almost coincide and their evolution in time is nicely fitted by the Monte Carlo estimated fractional moment (for $\delta=0.6$). On the contrary, when $\beta$ is large ($\beta=1.8$, Figure 7), the Monte Carlo estimate (dots) nicely fits curve (15), while the analytical curve (14) is neatly different from the other two (for $\delta=1.2$).

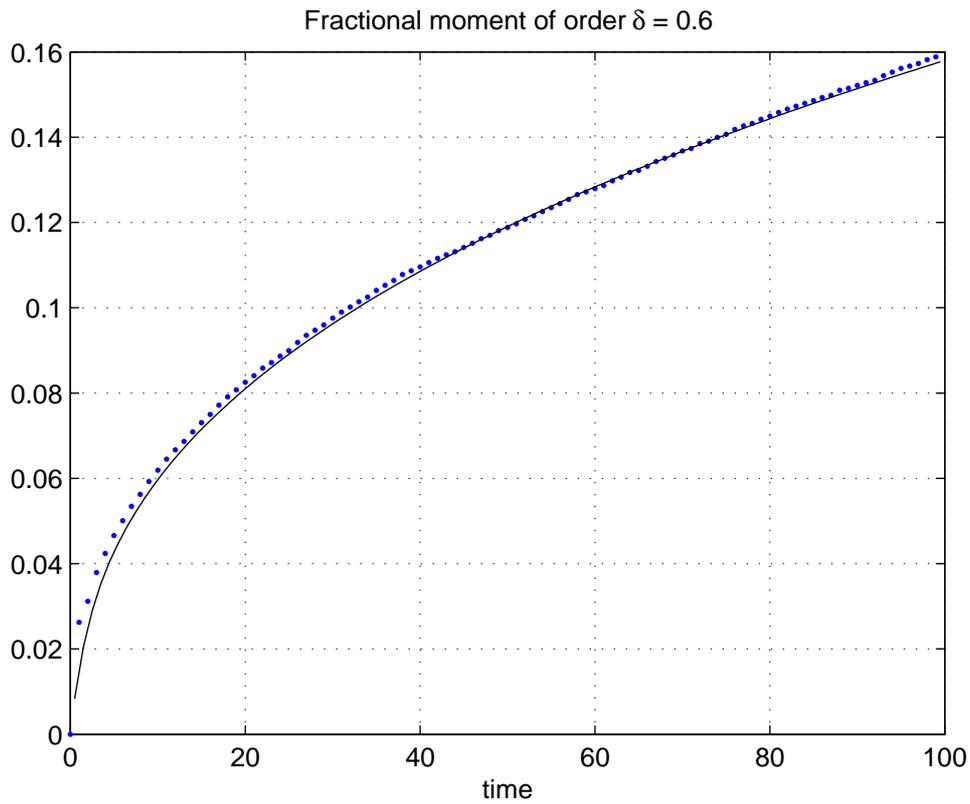

*Figure 6. Monte Carlo, numerically computed curve (15) and analytical expression (14) almost coincide.*

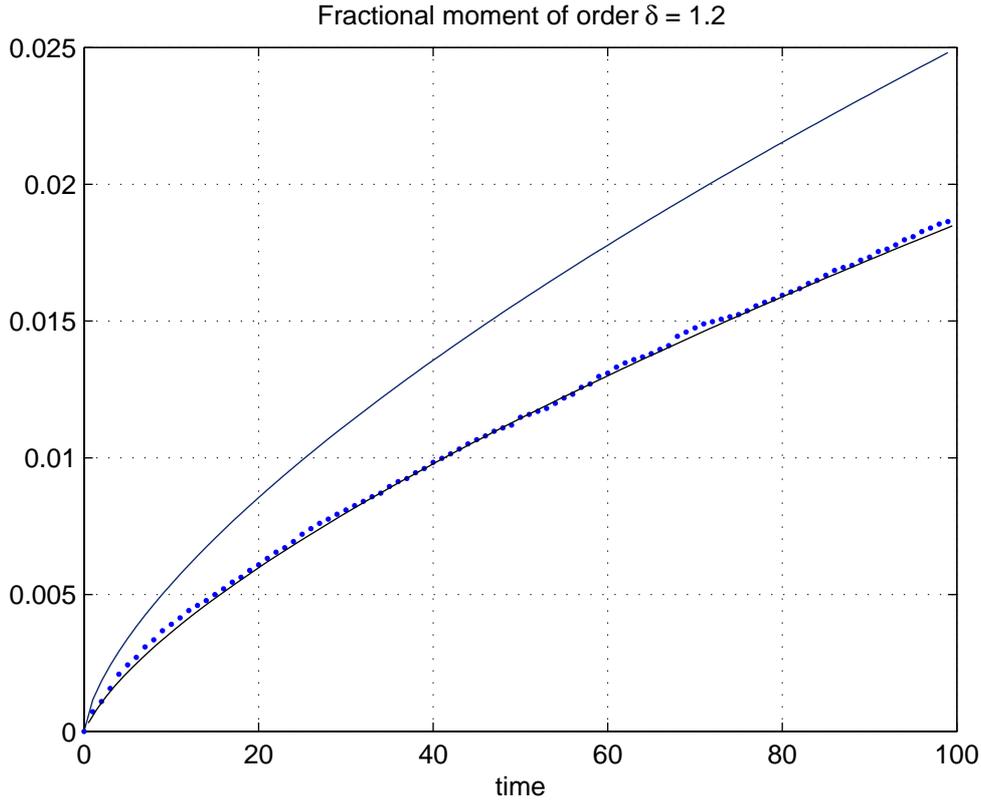

*Figure 7. Monte Carlo (dots) nicely fits curve (15), while expression (14) lies neatly above the others.*

## 6. Conclusions

Subsurface particle transport is commonly described by means of the standard advection-dispersion equation (ADE), which relies on a "Fickian assumption": the Central Limit Theorem is assumed to hold true for the particle ensemble, so that its spread will grow linearly in time. However, when transport phenomena are such that the tracers flow along preferential "streams" through the surrounding environment, the Gaussian hypothesis doesn't hold anymore and – at least for a transient phase – anomalous diffusion should be taken into account in order to suitably interpret and describe the "non-Fickian" behaviour of the particle plume. Evidences of superdiffusive behavior have been recently reported in laboratory-scale experiments as well as in direct field measurements of slowly-decaying radioactive particles transport: the contaminant spread is observed to grow more than linearly in time. This topic is of utmost interest e.g. for the dimensioning of nuclear waste repositories, since the adoption of Fickian advection-dispersion models would lead to an underestimation of the effective contaminant spread in the geological structures. Several approaches have been proposed to interpret and model superdiffusive transport: one of the most successful and most widely adopted is the Continuous-Time Random Walk approach, based on the Chapman-Kolmogorov probability balance. Within the CTRW, each superdiffusive particle (walker) is supposed to perform random jumps (sampled from a power law distribution of exponent $\beta$, ranging from *1* to *2*) separated by random waiting times (sampled by any finite mean distribution, such as an exponential pdf), during which the walker stays at rest in the previously reached position.

This Markovian stochastic process – also called "Lévy flight" – gives rise to a distribution $P(x,t)$ of being at position *x* at time *t*, whose variance grows more than linearly in time. In general, no analytical solution is available for $P(x,t)$ within the CTRW approach. However, in the Fourier and Laplace transformed space, $P(k,u)$ has a simple algebraic expression. A truncation to the first non-constant term in the Fourier space gives rise to the so-called FDE model, which admits an elegant analytical solution in terms of the Fox's *H* function in the direct space. In order to quantitatively evaluate the relevance of this approximation in the transformed space, a Monte Carlo solution of the CTRW model has been here computed by simulating an ensemble of particles wandering in a medium according to a superdiffusive behavior, with a given $\beta$. It turns out that when the characteristic exponent $\beta$ of the jump lengths distribution is small (in the allowed range, i.e. approximately $\beta \leq 1.5$) the FDE model nicely fits the Monte Carlo curve, consequently yielding accurate solutions for the superdiffusive transport. On the contrary, as $\beta$ approaches 2, the Monte Carlo neatly disagrees with the analytical FDE solution, showing that the approximations therein introduced are no more negligible.

To proceed further, we notice that the FDE model considers only a contribution of order $\beta$ in the asymptotic expansion of the Fourier transform of the jump lengths distribution and neglects a successive quadratic term. This latter is effectively of scarce importance when $\beta$ is small, in the so-called "diffusion limit", i.e. for large $x$. When, on the contrary, $\beta$ approaches the superior limit 2, which corresponds to a Fickian diffusive behavior for the walkers, the contribution of the quadratic term becomes comparable to the one of the characteristic exponent power: therefore, we expect the exact solutions to be increasingly different from those of the approximated FDE model.

In this paper, we have proposed an analytical expression for the Fourier transform of the jump lengths distributions which takes into account also the quadratic term: correspondingly, we have obtained an hybrid solution $P(x,t)$ of the CTRW model (partly analytical and partly numerical), which turned out to be in very good agreement with the results of the Monte Carlo simulation. The same idea has been extended to the fractional order moments of the $P(x,t)$ distribution: again, it turns out that for small values of the characteristic exponent the Monte Carlo is in good agreement with the analytical expression for the fractional moments, which can be derived explicitly by exploiting the properties of the $H$ function within the FDE model. On the contrary, when the characteristic exponent approaches 2, the Monte Carlo disagrees with this analytical expression and agrees with the hybrid solution. Both the comparisons here performed confirm that the FDE yields accurate solutions only when the characteristic exponent is small, evidencing the need to improve FDE by taking into account the quadratic term contribution. Moreover, Monte Carlo, thanks to its flexibility, may be considered as a very effective way to validate the accuracy of approximate solutions of the CTRW model. Parameterization of the superdiffusive model is made possible by comparing e.g. an experimental breakthrough concentration profile with the corresponding analytical curve as obtained from the anomalous diffusion model (which is a function of the main parameter $\beta$). This would in principle allow to quantify $\beta$ and to determine how far the particle flow is from the usual Fickian behaviour (i.e. $\beta=2$). By taking into account the previous considerations, we would be led to the conclusion that when the diffusion regime of the superdiffusive particle plume is nearly Fickian the improved FDE (10*) we have proposed would give a better estimate of the exponent $\beta$.

**Appendix:** Fourier transform of pdf (4)

We begin by considering the Fourier integral $\int_{-\infty}^{+\infty} |x|^{-1-\beta} e^{-ikx} dx$. By taking into account the symmetry of the function and recalling that it is defined in the domain $|x| \geq \sigma$, we can resort to the cosine transform:

$$\int_{-\infty}^{+\infty} |x|^{-1-\beta} e^{-ikx} dx = 2 \int_{\sigma}^{+\infty} x^{-1-\beta} \cos(kx) \, dx \, .$$

To evaluate this integral we resort to the temporary introduction of an exponential function of parameter $\alpha$, so that

$$2 \int_{\sigma}^{+\infty} x^{-1-\beta} \cos(kx) \, dx = 2 \lim_{\alpha \to 0} \int_{\sigma}^{+\infty} x^{-1-\beta} \cos(kx) e^{-\alpha x} dx$$

Hence, we integrate thrice by parts, neglecting all the terms where the parameter $\alpha$ directly appears at the numerator, as they will be later cancelled by passing to the limit:

$$2 \int_{\sigma}^{+\infty} x^{-1-\beta} \cos(kx) e^{-\alpha x} dx = -\frac{2\cos(kx)}{\beta} x^{-\beta} e^{-\alpha x} \Big|_{\sigma}^{+\infty} + \frac{2}{\beta} \int_{\sigma}^{+\infty} x^{-\beta} \left( -k \sin(kx) e^{-\alpha x} - \alpha \cos(kx) e^{-\alpha x} \right) dx =$$

$$= \frac{2\cos(k\sigma)}{\beta} \sigma^{-\beta} e^{-\alpha \sigma} - \frac{2k}{\beta} \left[ \frac{\sin(kx)}{1-\beta} x^{1-\beta} e^{-\alpha x} \Big|_{\sigma}^{+\infty} - \frac{1}{1-\beta} \int_{\sigma}^{+\infty} x^{1-\beta} \left( k \cos(kx) e^{-\alpha x} - \alpha \sin(kx) e^{-\alpha x} \right) dx \right] =$$

$$= \frac{2\sigma^{-\beta} \cos(k\sigma) e^{-\alpha \sigma}}{\beta} + \frac{2k \sigma^{1-\beta} \sin(k\sigma) e^{-\alpha \sigma}}{\beta(1-\beta)} +$$

$$+ \frac{2k^2}{\beta(1-\beta)} \left[ \frac{1}{2-\beta} x^{2-\beta} \cos(kx) e^{-\alpha x} \Big|_{\sigma}^{+\infty} - \int_{\sigma}^{+\infty} \frac{x^{2-\beta}}{2-\beta} \left( -k \sin(kx) e^{-\alpha x} - \alpha \cos(kx) e^{-\alpha x} \right) dx \right] =$$

$$= \frac{2\sigma^{-\beta}\cos(k\sigma)e^{-\alpha\sigma}}{\beta} + \frac{2k\sigma^{1-\beta}\sin(k\sigma)e^{-\alpha\sigma}}{\beta(1-\beta)} - \frac{2k^2\sigma^{2-\beta}\cos(k\sigma)e^{-\alpha\sigma}}{\beta(1-\beta)(2-\beta)} + \frac{2k^3}{\beta(1-\beta)(2-\beta)}\int_\sigma^{+\infty} x^{2-\beta}\left(k\sin(kx)e^{-\alpha x}\right)dx$$

By putting $z=kx$ in the last integral we have: $\int_\sigma^{+\infty} x^{2-\beta}\sin(kx)e^{-\alpha x}dx = k^{\beta-3}\int_{k\sigma}^{+\infty} z^{2-\beta}\sin(z)e^{-\frac{\alpha}{k}z}dz$.

Recalling that $\lambda(k) = \frac{\beta\sigma^\beta}{2}\int_{-\infty}^{+\infty}|x|^{-1-\beta}e^{-ikx}dx$ and passing to the limit $\alpha \to 0$ we get:

$$\lambda(k) = \cos(k\sigma) + \frac{k\sigma\sin(k\sigma)}{(1-\beta)} - \frac{k^2\sigma^2\cos(k\sigma)}{(1-\beta)(2-\beta)} + \lim_{\alpha\to 0}\frac{(|k|\sigma)^\beta}{(1-\beta)(2-\beta)}\int_{k\sigma}^{+\infty} z^{2-\beta}\sin(z)e^{-\frac{\alpha}{k}z}dz$$

We consider now the expansion for small $k$: $\sin(k\sigma) = k\sigma + o(k^3)$, $\cos(k\sigma) = 1 - \frac{(k\sigma)^2}{2} + o(k^4)$ and $\int_{k\sigma}^{+\infty}.\approx \int_0^{+\infty}..$

Then,

$$\lambda(k) = 1 - \frac{(k\sigma)^2}{2} + \frac{(k\sigma)^2}{(1-\beta)} - \frac{(k\sigma)^2}{(1-\beta)(2-\beta)} + \lim_{\alpha\to 0}\frac{(|k|\sigma)^\beta}{(1-\beta)(2-\beta)}\int_0^{+\infty} z^{2-\beta}\sin(z)e^{-\frac{\alpha}{k}z}dz + o(k^4).$$

By resorting to the noteworthy relationship [38]:

$$\int_0^{+\infty} x^{\mu-1}e^{-\gamma x}\sin(\delta x)dx = \frac{\Gamma(\mu)}{(\gamma^2+\delta^2)^{\frac{\mu}{2}}}\sin\left(\mu\tan^{-1}\left(\frac{\delta}{\gamma}\right)\right)$$

and identifying the parameters $\mu-1 = 2-\beta$, $\delta = 1$ and $\gamma = \frac{\alpha}{k}$, we have

$$\lambda(k) = 1 - \frac{(k\sigma)^2}{2} + \frac{(k\sigma)^2}{(1-\beta)} - \frac{(k\sigma)^2}{(1-\beta)(2-\beta)} + \lim_{\alpha\to 0}\frac{(|k|\sigma)^\beta}{(1-\beta)(2-\beta)}\frac{\Gamma(3-\beta)}{\left(\left(\frac{\alpha}{k}\right)^2+1\right)^{\frac{3-\beta}{2}}}\sin\left((3-\beta)\tan^{-1}\left(\frac{k}{\alpha}\right)\right) + o(k^4)$$

Then, explicitly solving the last limit and suitably simplifying, we finally obtain:

$$\lambda(k) = 1 - \left[\Gamma(1-\beta)\cos\left(\frac{\pi}{2}\beta\right)\right](|k|\sigma)^\beta + \left[\frac{\beta}{2(2-\beta)}\right](k\sigma)^2 + o(k^4).$$